\documentclass[preprint,showpacs,amsmath,amssymb]{revtex4}
\usepackage{mathrsfs}

\usepackage{graphicx,color}
\usepackage{dcolumn}
\usepackage{bm}
\usepackage{CJK}

\begin{document}
\begin{CJK*}{GBK}{song}

\title{Nuclear $\beta^+$/EC decays in covariant density functional theory and the impact of isoscalar proton-neutron pairing}

\author{Z. M. Niu $^1$}
\author{Y. F. Niu $^2$}
\author{Q. Liu $^1$}
\author{H. Z. Liang $^{3,4}$}\email{haozhao.liang@riken.jp}
\author{J. Y. Guo $^1$}

\affiliation{$^1$School of Physics and Material Science, Anhui University,
             Hefei 230039, China}
\affiliation{$^2$Institute of Fluid Physics, China Academy of Engineering
             Physics, Mianyang 621900, China}
\affiliation{$^3$RIKEN Nishina Center,
             Wako 351-0198, Japan}
\affiliation{$^4$State Key Laboratory of Nuclear Physics and Technology,
             School of Physics, Peking University, Beijing 100871, China}

\date{\today}

\begin{abstract}
Self-consistent proton-neutron quasiparticle random phase approximation
based on the spherical nonlinear point-coupling relativistic
Hartree-Bogoliubov theory is established and used to investigate the
$\beta^+$/EC-decay half-lives of neutron-deficient Ar, Ca, Ti, Fe, Ni,
Zn, Cd, and Sn isotopes. The isoscalar proton-neutron pairing is found
to play an important role in reducing the decay half-lives, which is
consistent with the same mechanism in the $\beta$ decays of neutron-rich
nuclei. The experimental $\beta^+$/EC-decay half-lives can be well
reproduced by a universal isoscalar proton-neutron pairing strength.
\end{abstract}

\maketitle

Nuclear $\beta$ decays play important roles in many subjects of nuclear
physics. Specifically, the investigation of $\beta$ decay provides
information on the spin and isospin dependence of the effective nuclear
interaction, as well as on nuclear properties such as
masses~\cite{Lunney2003RMP}, shapes~\cite{Nacher2004PRL}, and energy
levels~\cite{Tripathi2008PRL}. Moreover, nuclear $\beta$ decays are also
important in nuclear astrophysics, because they set the time scale of the
rapid neutron-capture process ($r$-process)~\cite{Burbidge1957RMP,
Langanke2003RMP, Qian2007PRp, Sun2008PRC, Niu2009PRC}, which is a major
mechanism for producing the elements heavier than iron. In addition,
nuclear $\beta$ decays can provide tests for the electroweak standard
model~\cite{Severijns2006RMP, Liang2009PRC, Hardy2010RPP}. With the
development of radioactive ion beam facilities, the measurement of nuclear
$\beta$-decay half-lives has achieved great progress in recent
years~\cite{Grevy2004PLB, Hosme2005PRL, Nishimura2011PRL, Audi2012CPC}.

On the theoretical side, apart from the macroscopic gross
theory~\cite{Takahashi1975ADNDT}, two different microscopic approaches
have been widely used to describe and predict the nuclear $\beta$-decay
rates. They are the shell model \cite{Langanke2003RMP} and the
proton-neutron quasiparticle random phase approximation
(QRPA)~\cite{Moller1997ADNDT, Hirsch1993ADNDT, Ni2012JPG}. While the shell
model takes into account the detailed structure of the $\beta$-strength
function, the proton-neutron QRPA approach provides a systematic
description of $\beta$-decay properties of arbitrarily heavy nuclei. In
order to reliably predict properties of thousands of unknown nuclei
relevant to the $r$-process, the self-consistent QRPA approach has become
a current trend in nuclear structure study, including those based on the
Skyrme-Hartree-Fock-Bogoliubov (SHFB) theory~\cite{Engel1999PRC} and the
covariant density functional theory (CDFT)~\cite{Niksic2005PRC,
Marketin2007PRC, Niu2012PLB}.

In the CDFT framework, the self-consistent proton-neutron RPA was first
developed based on the meson-exchange relativistic Hartree (RH)
approach~\cite{Conti1998PLB}. To describe the spin-isospin excitations in
open shell nuclei, it has been extended to the QRPA based on the
relativistic Hartree-Bogoliubov (RHB) approach~\cite{Paar2004PRC} and
employed to calculate the $\beta$-decay half-lives of neutron-rich nuclei
in the $N\approx 50$ and $N\approx 82$ regions~\cite{Niksic2005PRC,
Marketin2007PRC}. In addition, based on the meson-exchange relativistic
Hartree-Fock (RHF) approach~\cite{Long2006PLB, Long2007PRC}, the
self-consistent proton-neutron RPA has been formulated~\cite{Liang2008PRL}
and well reproduces the spin-isospin excitations in doubly magic nuclei,
without any readjustment of the parameters of the covariant energy density
functional~\cite{Liang2008PRL, Liang2012PRC}. Recently, the
self-consistent QRPA based on the relativistic Hartree-Fock-Bogoliubov
(RHFB) approach~\cite{Long2010PRCa, Long2010PRCb} was developed and a
systematic study on the $\beta$-decay half-lives of neutron-rich even-even
nuclei with $20 \leqslant Z \leqslant 50$ has been
performed~\cite{Niu2012PLB}. Similar to the non-relativistic
calculations~\cite{Engel1999PRC}, it is found that the isoscalar ($T=0$)
proton-neutron pairing plays a very important role in reducing the decay
half-lives. In particular, with an isospin-dependent $T=0$ proton-neutron
pairing interaction as a function of $N-Z$, available data in the whole
region of $20 \leqslant Z \leqslant 50$ can be well
reproduced~\cite{Niu2012PLB}. So far, these self-consistent investigations
mainly focus on the neutron-rich side.

During the past years, the CDFT framework has been reinterpreted by the
relativistic Kohn-Sham scheme, and the functionals have been developed
based on the zero-range point-coupling interactions~\cite{Nikolaus1992}.
In this framework, the meson exchange in each channel is replaced by the
corresponding local four-point contact interaction between nucleons. Such
point-coupling model has attracted more and more attentions due to its
simplicity and several other advantages~\cite{Niksic2011PPNP}. For
example, it is even possible to include the effects of Fock terms in a
local RHF equivalent scheme~\cite{Liang2012PRCb,Gu2013PRC}. With either
nonlinear or density-dependent effective interactions, the point-coupling
models have achieved satisfactory descriptions for infinite nuclear matter
and finite nuclei on a level of accuracy comparable to that of
meson-exchange models~\cite{Burvenich2002PRC, Niksic2008PRC}. Recently, a
new nonlinear point-coupling effective interaction
PC-PK1~\cite{Zhao2010PRC} was proposed, which well reproduces the
properties of nuclear matter and finite nuclei including the ground-state
and low-lying excited states~\cite{Zhao2010PRC, Hua2012SCMPA, Mei2012PRC}.
In particular, the PC-PK1 provides a good isospin dependence of binding
energy along either the isotopic or the isotonic chain, which makes it
reliable for the applications in exotic nuclei~\cite{Zhao2010PRC,
Hua2012SCMPA}. Based on the point-coupling effective Lagrangian, the
spherical (Q)RPA in non-charge-exchange channel has been formulated and
well reproduces the excitation energies of giant
resonances~\cite{Niksic2005PRCa, Niksic2008PRC,Liang2013arXiv}.

In this work, the self-consistent proton-neutron QRPA based on the
spherical nonlinear point-coupling relativistic Hartree-Bogoliubov theory
is established. This newly developed approach will be used to investigate
the $\beta^+$/EC decays in neutron-deficient isotopes around the proton
magic numbers $Z=20$, $28$, and $50$ with the PC-PK1 effective
interaction. Special attention will be paid to the effects of the $T=0$
proton-neutron pairing on the decay half-lives.

For a self-consistent QRPA calculation, the particle-hole (p-h) and
particle-particle (p-p) residual interactions should be derived from the
same energy density functional as ground state. Here we only collect the
essential expressions and refer the readers to Refs.~\cite{Paar2004PRC,
Finelli2007NPA} for some details of the relativistic proton-neutron QRPA.

For the p-h residual interaction, only the isovector channel of the
effective interaction contributes to the charge-exchange excitations. The
isovector-vector (TV) interaction in the present relativistic
point-coupling model reads
\begin{eqnarray}\label{Eq:TVph}
  V_{TV}(1,2) = (\alpha_{TV}+\delta_{TV}\Delta)
                [\gamma_0\gamma^\mu\vec\tau]_1
                [\gamma_0\gamma_\mu\vec\tau]_2
                \delta(\boldsymbol{r}_1 - \boldsymbol{r}_2).
\end{eqnarray}
Similar to Refs.~\cite{Niksic2005PRC, Finelli2007NPA}, although the direct
one-pion contribution is absent in the ground-state description under the
Hartree approximation, it has to be included in the calculation of
spin-isospin excitations. The corresponding interaction reads
\begin{equation}\label{Eq:piph}
  V_\pi(1,2) = -\frac{f_\pi^2}{m_\pi^2}
                [\vec\tau\gamma_0\gamma_5\gamma^k\partial_k]_1
                [\vec\tau\gamma_0\gamma_5\gamma^l\partial_l]_2
                D_\pi(1,2),
\end{equation}
where $m_\pi = 138.0~\textrm{MeV}$ and $f_\pi^2 / 4\pi = 0.08$, while
$D_\pi(1,2)$ denotes the finite-range Yukawa type propagator. The
derivative type of the pion-nucleon coupling necessitates the inclusion of
the zero-range counter term, which accounts for the contact part of the
pion-nucleon interaction
\begin{eqnarray}\label{Eq:deltapiph}
    V_{\delta \pi}(1,2)
  = g' \frac{f_\pi^2}{m_\pi^2}
       [\vec{\tau}\gamma_0\gamma_5\boldsymbol{\gamma}]_1
       [\vec{\tau}\gamma_0\gamma_5\boldsymbol{\gamma}]_2
       \delta(\boldsymbol{r}_1 - \boldsymbol{r}_2),
\end{eqnarray}
where the $g'$ is adjusted to reproduce the excitation energy of the
Gamow-Teller (GT) resonances in $^{208}$Pb. For the effective interaction
PC-PK1, $g'$ is determined to be $0.52$.

For the p-p residual interaction, we employ the pairing part of the Gogny
force for the isovector ($T = 1$) proton-neutron pairing interaction,
\begin{eqnarray}\label{Eq:T1}
        V_{T=1}(1,2)
   &=&  \sum_{i=1,2} e^{-[(\boldsymbol{r}_1-\boldsymbol{r}_2)/\mu_i]^2}\nonumber\\
   & &  (W_i + B_i P^\sigma - H_i P^\tau -M_i P^\sigma P^\tau),
\end{eqnarray}
with the parameter set D1S~\cite{Berger1984NPA} for $\mu_i, W_i, B_i,
H_i$, and $M_i$. For the isoscalar ($T = 0$) proton-neutron pairing
interaction in the QRPA calculation, we employ a similar interaction as in
Refs.~\cite{Engel1999PRC, Niksic2005PRC, Marketin2007PRC, Niu2012PLB}:
\begin{eqnarray}\label{Eq:T0}
    V_{T=0}(1,2)=-V_0 \sum_{i=1,2} g_i e^{-[(\boldsymbol{r}_1-\boldsymbol{r}_2)/\mu_i]^2}
                 \hat{\prod}_{S=1,T=0},
\end{eqnarray}
with $\mu_1=1.2$ fm, $\mu_2=0.7$ fm, $g_1=1$, $g_2=-2$. The operator
$\hat{\prod}_{S=1,T=0}$ projects onto states with $S=1$ and $T=0$. The
strength parameter $V_0$ is determined by fitting to known half-lives.

Similar as in Refs.~\cite{Gove1971NDT, Sarriguren2005PRC, Moreno2006PRC},
the $\beta$-decay half-life of an even-even nucleus is calculated in the
allowed approximation with
\begin{eqnarray}\label{Eq:BetaDecayRate}
    T_{1/2}
  =\frac{D}
        {\sum_\nu [  (g_A/g_V)_{\textrm{eff}}^2 B_{\textrm{GT}}(E_\nu)
                   + B_{\textrm{F}}(E_\nu) ]
                  f(Z,E_\nu)},
\end{eqnarray}
where $D=6163.4$ s and $(g_A/g_V)_{\textrm{eff}}=1$ is the effective ratio
of axial and vector coupling constants. $B_{\textrm{F}}(E_\nu)$ and
$B_{\textrm{GT}}(E_\nu)$ are the transition probabilities for allowed
Fermi (F) and GT transitions, which are calculated from the QRPA approach.
In $\beta^+/\textrm{EC}$ decay of neutron-deficient nucleus, $f(Z,E_\nu)$
consists of two parts, positron emission ($f^{\beta^+}$) and electron
capture ($f^{\textrm{EC}}$). The Fermi integral for positron emission
$f^{\beta^+}(Z,E_m)$ is given by
\begin{eqnarray}
  f^{\beta^+}(Z,E_m) =
               \int_{m_e}^{E_m}
               p_e E_e (E_m-E_e)^2 F_0(Z,E_e)dE_e,
\end{eqnarray}
where $p_e$ and $E_e$ are the emitted electron momentum and energy,
respectively. $F_0(Z,E_e)$ is the Fermi function including Coulomb
screening and relativistic nuclear finite-size
corrections~\cite{Langanke2003RMP}. In the self-consistent QRPA approach,
the $\beta^+$-decay energy $E_m$, i.e., the energy difference between the
initial and final states, can be calculated using the QRPA:
\begin{equation}\label{Eq:BetaDecayEm}
    E_m = -\Delta_{nH} - m_e - E_{\textrm{QRPA}},
\end{equation}
where $E_{\textrm{QRPA}}$ is the QRPA energy with respect to the
ground-state of the parent nucleus and corrected by the difference of the
proton and neutron Fermi energies in the parent nucleus \cite{Niu2012PLB},
(i.e., $E_{\textrm{QRPA}} - (\lambda_p - \lambda_n)$ with the definitions
in Ref.~\cite{Engel1999PRC}), $m_e$ and $\Delta_{nH}$ are the positron
mass and the mass difference between the neutron and the hydrogen atom,
respectively. Because the emitted positron energy must be higher than its
rest mass, the final states must be those with excitation energies
$E_{\textrm{QRPA}}< -\Delta_{nH} - 2m_e$. Moreover, the decay function
$f^{\textrm{EC}}$ for electron capture has also been included following
Ref.~\cite{Moreno2006PRC}:
\begin{equation}\label{Eq:BetaDecayEm}
    f^{\textrm{EC}} = \frac{\pi}{2}\sum_x q_x^2 g_x^2 B_x,
\end{equation}
where $x$ denotes the atomic subshell from which the electron is captured,
$q$ is the neutrino energy, $g$ is the radial component of the bound-state
electron wave function at the nuclear surface, and $B$ stands for other
exchange and overlap corrections. The energy threshold for EC is $2m_e$
higher than the $\beta^+$ decay, i.e., $E_{\textrm{QRPA}}<-\Delta_{nH}$.

\begin{figure}
  \includegraphics[width=6cm]{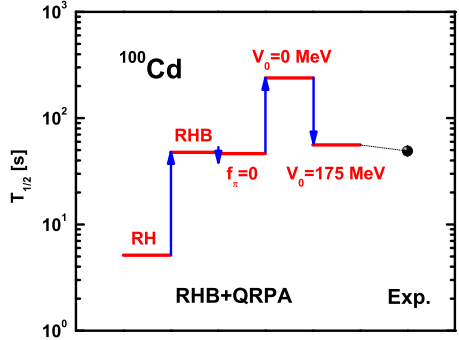}\\
  \caption{(Color online) The $\beta^+/\textrm{EC}$-decay half-life of $^{100}$Cd
  calculated by the self-consistent RHB + QRPA approach with the effective interaction PC-PK1~\cite{Zhao2010PRC}
  without and with the $T = 0$ proton-neutron pairing.
  The unperturbed results obtained by the RH and RHB approaches, and the
  QRPA result excluding the pion-nucleon p-h residual interactions are denoted by
  RH, RHB, and $f_\pi = 0$, respectively. For comparison, the experimental
  value~\cite{Audi2012CPC} is also shown.}
  \label{fig1}
\end{figure}

We first focus on the $\beta^+/\textrm{EC}$-decay half-life of $^{100}$Cd,
and show the corresponding contributions of p-h and p-p interactions in
Fig.~\ref{fig1}. By comparing the unperturbed results obtained by the RH
and RHB approaches, it is clear that the $T = 1$ proton-neutron pairing
interaction in Eq.~(\ref{Eq:T1}) plays an important role in the ground
state for the half-life calculation. Note that its corresponding p-p
residual interaction is not included in the QRPA for the unnatural parity
modes. Then, the TV p-h residual interaction in Eq.~(\ref{Eq:TVph}) is
introduced based on the RHB unperturbed result, however, its influence on
the half-life calculation is almost negligible. Furthermore, the half-life
substantially increases when the pion-nucleon interaction in
Eq.~(\ref{Eq:piph}) and its zero-range counter term in
Eq.~(\ref{Eq:deltapiph}) are included, because their total contributions
are repulsive and dominant in p-h residual interactions for the GT
excitations. Finally, it is found that the calculated half-lives are very
sensitive to the $T = 0$ proton-neutron pairing interaction in
Eq.~(\ref{Eq:T0}) by comparing the results with and without such p-p
residual interaction. In previous studies~\cite{Engel1999PRC,
Niksic2005PRC, Marketin2007PRC, Niu2012PLB}, the strength $V_0$ is usually
determined by adjusting QRPA results to empirical half-lives. In this
work, we take $^{100}$Cd as the reference nucleus, and the value of $V_0$
is determined to be $175$ MeV.

\begin{figure}
  \includegraphics[width=6cm]{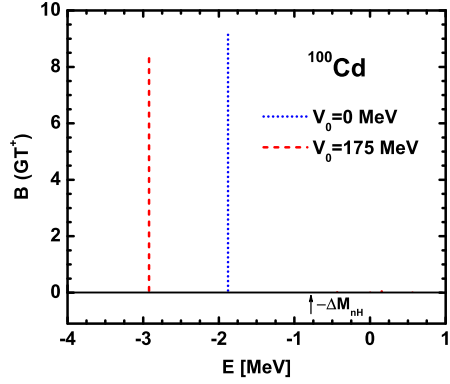}\\
  \caption{(Color online) Gamow-Teller transition probabilities of
  $^{100}$Cd calculated by RHB + QRPA approach with the effective interaction
  PC-PK1 without and with the $T = 0$ proton-neutron pairing.
  The threshold for EC decay is shown with an arrow.}\label{fig2}
\end{figure}

As shown in Eq. (\ref{Eq:BetaDecayRate}), the nuclear $\beta$-decay
half-life is determined by the transition strength as well as the
transition energy which decides the value of $f(Z,E_\nu)$. In order to
illustrate the mechanism of the influence from $T = 0$ pairing on the
$\beta^+/\textrm{EC}$-decay half-life, the Gamow-Teller transition
strength distributions of $^{100}$Cd are shown in Fig.~\ref{fig2}. It is
seen that without $T = 0$ pairing there is mainly one transition with
$E_{\textrm{QRPA}}=-1.879$ MeV contributing to nuclear
$\beta^+/\textrm{EC}$ decay. This transition is dominated by the spin-flip
configuration $\pi 1g 9/2 \rightarrow \nu 1g 7/2$. Because both $\pi 1g
9/2$ and $\nu 1g 7/2$ orbitals are partially occupied, (the occupation
probabilities of $\pi 1g 9/2$ and $\nu 1g 7/2$ orbitals are $0.808$ and
$0.123$, respectively), the $T = 0$ pairing can substantially contribute
to the QRPA matrices related to the $\pi 1g 9/2 \rightarrow \nu 1g 7/2$
pair. When the attractive $T = 0$ pairing is included, the transition
built from the such configuration is lowered in energy, and thus the value
of function $f(Z,E_\nu)$ increases, while the GT strength decreases
slightly. As a result, the $\beta^+/\textrm{EC}$-decay half-life is
remarkably reduced.

\begin{figure}
  \includegraphics[width=8cm]{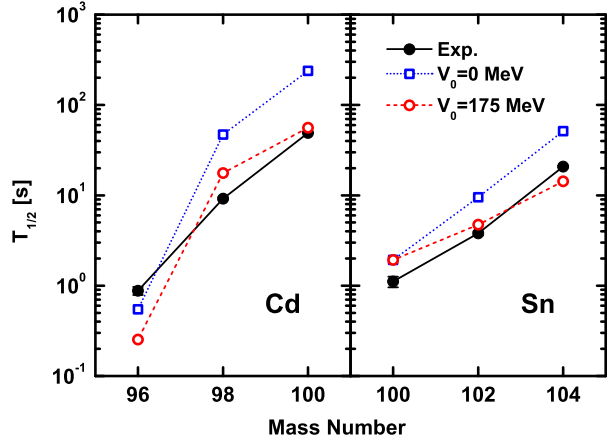}\\
  \caption{(Color online) Comparison of the calculated half-lives of Cd and
  Sn isotopes using RHB + QRPA approach and the effective interaction PC-PK1
  with experimental data~\cite{Audi2012CPC} (filled circles). The open squares and open circles
  denote the half-lives with the strength of $T=0$ pairing $V_0=0$ and $175$
  MeV, respectively.}
  \label{fig3}
\end{figure}

In order to further investigate the impact of $T=0$ pairing interaction on
the $\beta^+/\textrm{EC}$ decays, the corresponding half-lives for Cd and
Sn isotopes calculated by the self-consistent RHB + QRPA approach with and
without the $T = 0$ pairing interaction are shown in Fig.~\ref{fig3}. It
is clear that the calculations without the $T = 0$ pairing interaction
generally overestimate experimental values. By including the $T = 0$
pairing interaction, the calculated half-lives are significantly reduced
and well reproduce half-lives of $^{98, 100}$Cd and $^{100, 102, 104}$Sn.
Since $^{96}$Cd is a deformed nucleus, the underestimation of half-life
may originate from the deformation effect as the deformation can spread
and hinder the low-energy tails of the GT strength
distributions~\cite{Sarriguren2010PRC}. This effect is not included in the
present calculations. Therefore, it will be interesting to include
deformation degrees of freedom into the self-consistent QRPA calculations
and study their effects on $\beta$-decay half-lives in the future.

\begin{figure}
  \includegraphics[width=8cm]{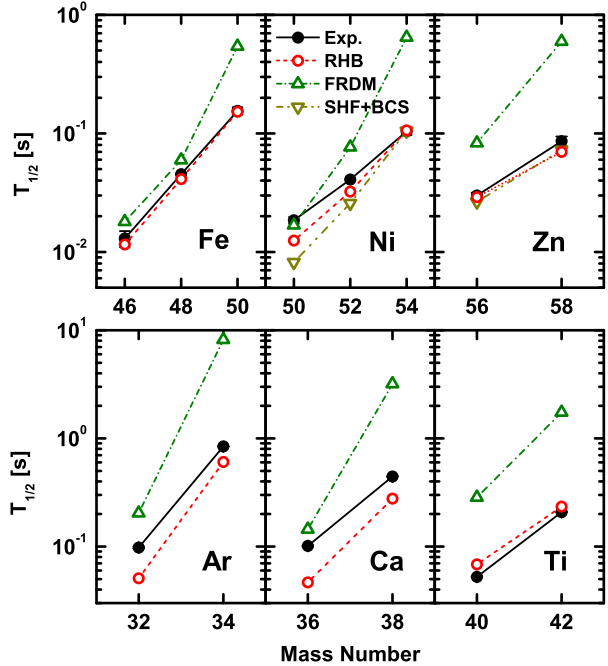}\\
  \caption{(Color online) Nuclear $\beta^+/\textrm{EC}$-decay half-lives for Fe, Ni, Zn,
  Ar, Ca, and Ti isotopes calculated by RHB + QRPA approach with the effective
  interaction PC-PK1 and $V_0=175$ MeV. For comparison, the experimental data~\cite{Audi2012CPC}
  (filled circles), as well as theoretical results obtained from
  FRDM + QRPA~\cite{Moller1997ADNDT} (open upward triangles) and
  SHF + BCS + QRPA~\cite{Sarriguren2011PRC} (open downward triangles)
  approaches are also shown.}
  \label{fig4}
\end{figure}

For the QRPA calculations~\cite{Niksic2005PRC, Marketin2007PRC}, the
strength of $T=0$ proton-neutron pairing $V_0$ was usually determined by
adjusting to the known half-life of selected nucleus in each isotopic
chain. However, very different values are found for neutron-rich nuclei of
different isotopic chains. Taking the Cd and Fe isotopic chains as
examples, the difference between the corresponding $V_0$ is about $100$
MeV~\cite{Niksic2005PRC, Marketin2007PRC}. This procedure, of course,
limits the prediction power of the model.

For improving this dilemma, an isospin-dependent form of $V_0$ was
proposed in Ref.~\cite{Niu2012PLB} and achieved great success in the
description of $\beta$-decay half-lives of neutron-rich nuclei with $20
\leqslant Z \leqslant 50$. In this isospin-dependent pairing strength, the
values of $V_0$ are nearly constant for nuclei with $N - Z < 5$, which is
exactly the case for the neutron-deficient nuclei in the same region,
i.e., $20 \lesssim Z \lesssim 50$. Therefore, we further calculate the
half-lives of Fe, Ni, Zn, Ar, Ca, and Ti isotopes with the same $V_0$
determined by the half-life of $^{100}$Cd. The results are shown in
Fig.~\ref{fig4}. For comparison, the calculated results obtained from the
macroscopic-microscopic finite-range droplet model (FRDM) +
QRPA~\cite{Moller1997ADNDT} and the SHF + BCS + QRPA with separable
residual interactions~\cite{Sarriguren2011PRC} are also shown.

It is found while these three approaches show similar isotopic trend of
nuclear half-lives, the present self-consistent RHB + QRPA calculations
reproduce the experimental data remarkably. In contrast, the SHF + BCS +
QRPA approach well reproduces the experimental half-life of $^{54}$Ni, but
it underestimates experimental half-lives of $^{50, 52}$Ni. For the FRDM +
QRPA approach, it almost systematically overestimates the experimental
half-lives. It has been pointed out that the overestimation of half-lives
in the FRDM + QRPA approach can be attributed partially to the neglect of
the $T = 0$ pairing~\cite{Engel1999PRC, Borzov2000PRC, Niu2012PLB}. This
is further supported by the present investigation on the $\beta^+$/EC
decays in neutron-deficient nuclei.

In summary, we have extended the self-consistent quasiparticle random
phase approximation approach to the charge-exchange channel based on the
relativistic Hartree-Bogoliubov model for the nonlinear point-coupling
effective interaction. This approach is then used to systematically
investigate the $\beta^+$/EC-decay half-lives of neutron-deficient nuclei
around the proton magic numbers $Z=20$, $28$, and $50$. It is found that
the calculated half-lives are very sensitive to the $T = 0$ proton-neutron
pairing interaction. By including the $T = 0$ pairing interaction, the
calculated half-lives are remarkably reduced, as the GT transitions are
substantially lowered in energy while the transition strengths only
slightly decrease. The experimental $\beta^+/\textrm{EC}$-decay half-lives
of Ar, Ca, Ti, Fe, Ni, Zn, Cd, and Sn isotopes can be well reproduced by a
universal $T = 0$ pairing strength.

This work was partly supported by the National Natural Science Foundation
of China under Grants No. 11205004, No. 11175001, and No. 11105006, the
211 Project of Anhui University under Grant No. 02303319-33190135, the
Talent Foundation of High Education of Anhui Province for Outstanding
Youth under Grant No. 2011SQRL014, and the Grant-in-Aid for JSPS Fellows
under Grant No. 24-02201.



\end{CJK*}
\end{document}